\documentclass[12pt]{book}

\usepackage[dvips]{graphicx,color}
\usepackage{makeidx,tsukuba}

\makeauthorindex
\makeindex

\begin{document}

\BookTitle{\itshape The 28th International Cosmic Ray Conference}
\CopyRight{\copyright 2003 by Universal Academy Press, Inc.}
\pagenumbering{arabic}

%
%    NOTE FOR THE EDITOR
%
%    Please, note that each \ref command in this text produces
%    a dot after the number.  For this reason I wrote everywhere
%    \ref{something}~ to simulate the double spacing used by TeX
%    after a full stop.  If you use better settings for the
%    automatic referencing, that do not produce the unwanted dot,
%    you have to replace every occurrence of "}~" with "}~".
%
%    This problem affects figure and table references, and it is
%    caused by the following definitions in `tsukuba.sty':
%    \def\thefigure{\@arabic\c@figure.}
%    \def\thetable{\@arabic\c@table.}
%

\chapter{The AMS-02 Time of Flight System.  Final Design}

\author{Diego Casadei,$^1$\\
 on behalf of the AMS TOF Collaboration: V.~Bindi,$^1$ N.~Carota,$^1$
 D.~Casadei,$^1$ G.~Castellini$^2$, F.~Cindolo,$^1$ A.~Contin,$^1$
 F.~Giovacchini,$^1$ P.~Giusti,$^1$ G.~Laurenti,$^1$ G.~Levi,$^1$
 R.~Martelli,$^1$ F.~Palmonari,$^1$ L.~Quadrani,$^1$ M.~Salvadore,$^1$
 C.~Sbarra,$^1$ and A.~Zichichi$^1$\\
{\it
(1) Bologna University and INFN Bologna, via Irnerio 46, I-40126
  Bologna, Italy\\
(2) CNR-IFAC, Via Panciatichi 64, I-50127 Firenze, Italy}
}%% end of author

\section*{Abstract}

 The AMS-02 detector is a superconducting magnetic spectrometer that
 will operate on the International Space Station. The time of flight
 (TOF) system of AMS-02 is composed by four scintillator planes with
 8, 8, 10, 8 counters each, read at both ends by a total of 144
 phototubes. This paper describes the new design, the expected
 performances, and shows preliminary results of the ion beam test
 carried on at CERN on October 2002.

\section{Introduction}

 The \emph{Alpha Magnetic Spectrometer} (AMS) is a particle detector
 that will be installed on the International Space Station (ISS) in
 2005 (NASA shuttle flight UF-4.1) to measure Cosmic Ray (CR) fluxes
 for at least three years.

 During the precursor flight aboard of the shuttle Discovery (NASA
 STS-91 mission, 2--12 June 1998), the test detector AMS-01 was
 operated for about 180 hours, collecting over one hundred millions CR
 events [1].  The AMS-02 detector, that will operate on ISS, is
 described in ref.~[2].  Scientific goals of AMS-02 are:
\begin{itemize}
  \item improved measurement of the antimatter fraction in cosmic
  rays;

  \item search for exotic particle annihilation signatures in the
  proton, electron and $\gamma$-ray spectra over the energy range
  1--$10^3$ GeV;

  \item high statistics measurements of the CR ion spectra below 1 TeV
  per nucleon.
\end{itemize}

\section{The time of flight system}

\begin{figure}[t]
  \centering
  \includegraphics[height=0.5\textwidth]{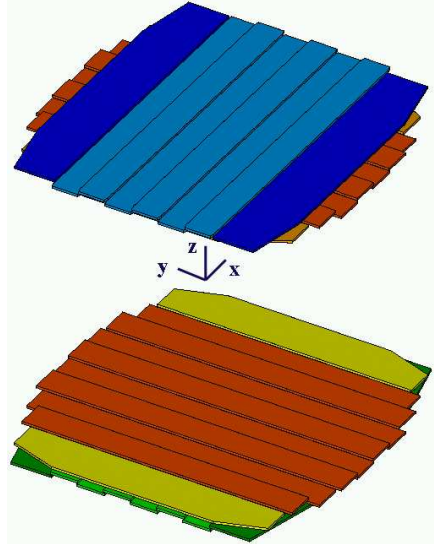}
  \caption{The scintillator paddles of the AMS-02 TOF system without
  light guides and phototubes.}\label{fig1}
\end{figure}

 The time of flight (TOF) system of AMS-02, developed in the
 laboratories of INFN Bologna, will be somewhat different from the TOF
 system of AMS-01 [3].  The whole system still consists of four
 plastic scintillator planes (two above and two below the magnet), but
 with 8, 8, 10, 8 counters each (instead of 14), as shown in
 figure~\ref{fig1}~ The counters of adjacent planes are orthogonal as
 in AMS-01, in order to provide a certain granularity at the trigger
 level, and in each plane they are overlapped by 5 mm.  Howewer, the
 outermost counters of AMS-02 have a trapezoidal shape in order to
 reduce the weight.

\begin{figure}[t]
  \centering
  \includegraphics[height=0.25\textwidth]{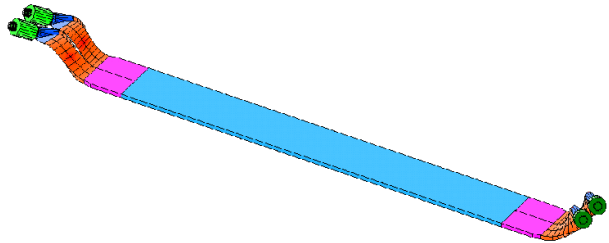}
  \caption{A complete TOF counter with 2 + 2 PMT and curved light
  guides.}\label{fig2}
\end{figure}

 All 1 cm thick scintillators are read at both ends by independently
 powered photo-multiplier tubes (PMT), connected with transparent
 light guides, as shown in figure~\ref{fig2}~ Due to the strong
 residual magnetic field in the PMT zones, the AMS-02 TOF counters
 adopt the Hamamatsu R5946 phototubes [4] and have curved light
 guides.  The total number of PMT is 144 and the optical contact with
 the light guides has been realized using soft transparent pads, that
 also provide a good mechanical coupling.  The same method was adopted
 for AMS-01 and proved to be able to survive both to the launch and to
 the landing of the shuttle Discovery without any damage to the PMT.

 The TOF system has the following essential tasks.  First, its signals
 are used to form the fast trigger (the very first level of the data
 acquisition chain).  Second, the particle time of flight is used to
 measured the velocity $\beta = v/c$ and to distinguish between upward
 and downward going particles.  This is fundamental in order to
 separate particles from antiparticles.  Third, the energy loss
 measurement is used by the TOF system to send to the trigger box a
 special flag for ions events, and to provide an independent particle
 charge measurement.

\section{The ion beam test preliminary results}

 Two AMS-02 TOF counters with $1 \times 12 \times 130$ cm$^3$
 polyvinyltoluene scintillators (Eljen EJ-200, ``C1'', and Bicron
 BC408, ``C2''), together with one AMS-01 counter used as reference,
 were tested at CERN on October 2002 with standard NIM and CAMAC
 electronics, on the ion beam provided by the SPS facility. The
 primary Pb beam was directed against 10--30 cm long Be targets,
 producing secondary particles and nuclei with charge spanning a very
 wide range: $Z = 1$--82.  The H8 selection line was tuned to obtain
 secondaries with $A/Z = 2$, 3/2, 7/4 and 1. For what concerns the
 scintillators, all primary particles were at their minimum ionization
 plateau (that is at Lorenz factors $>$ 3).

\begin{figure}[t]
\begin{minipage}{0.5\textwidth}
\includegraphics[height=0.9\linewidth]{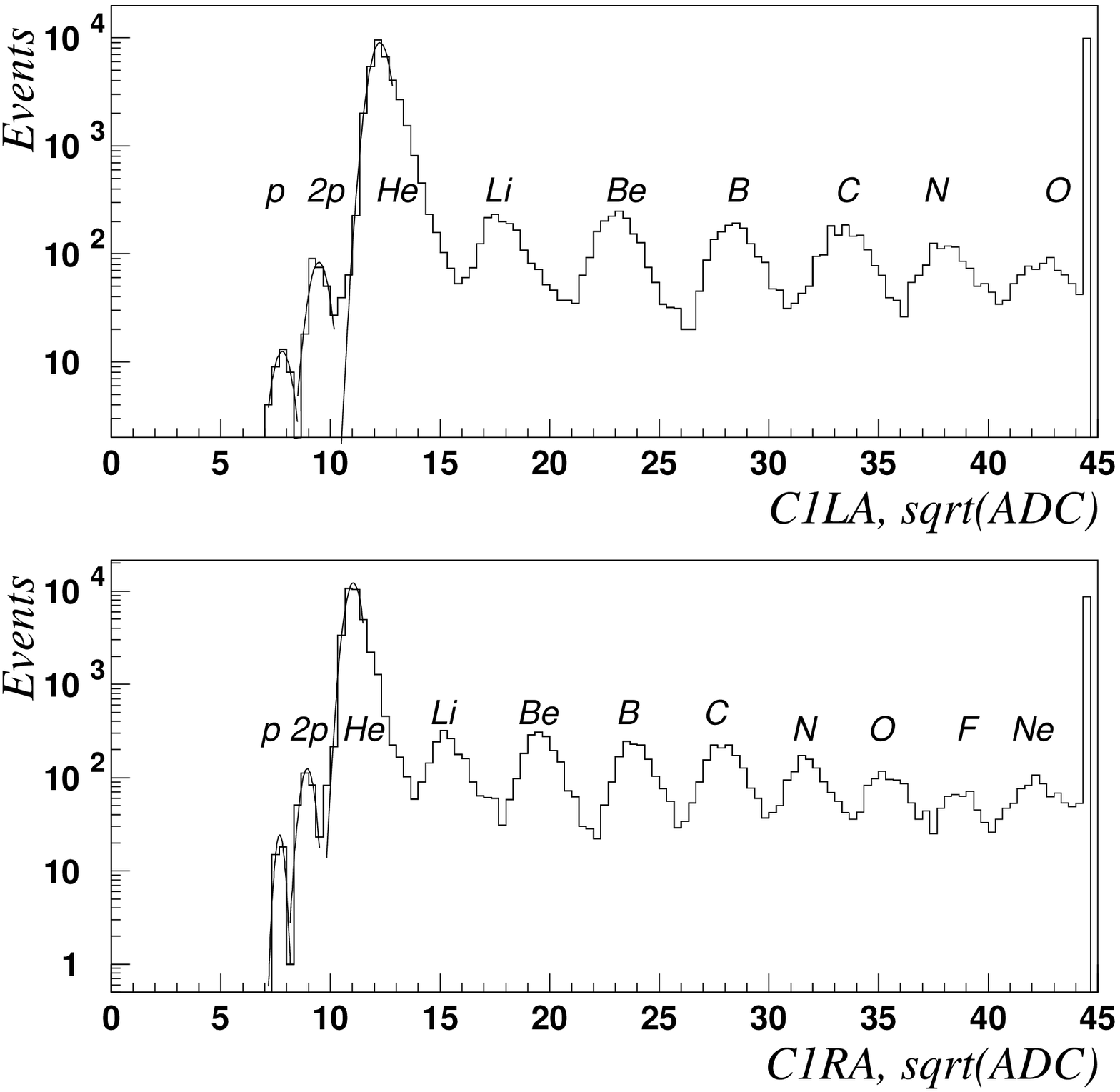}
\end{minipage}%
\begin{minipage}{0.5\textwidth}
~\hfill
\includegraphics[height=0.9\linewidth]{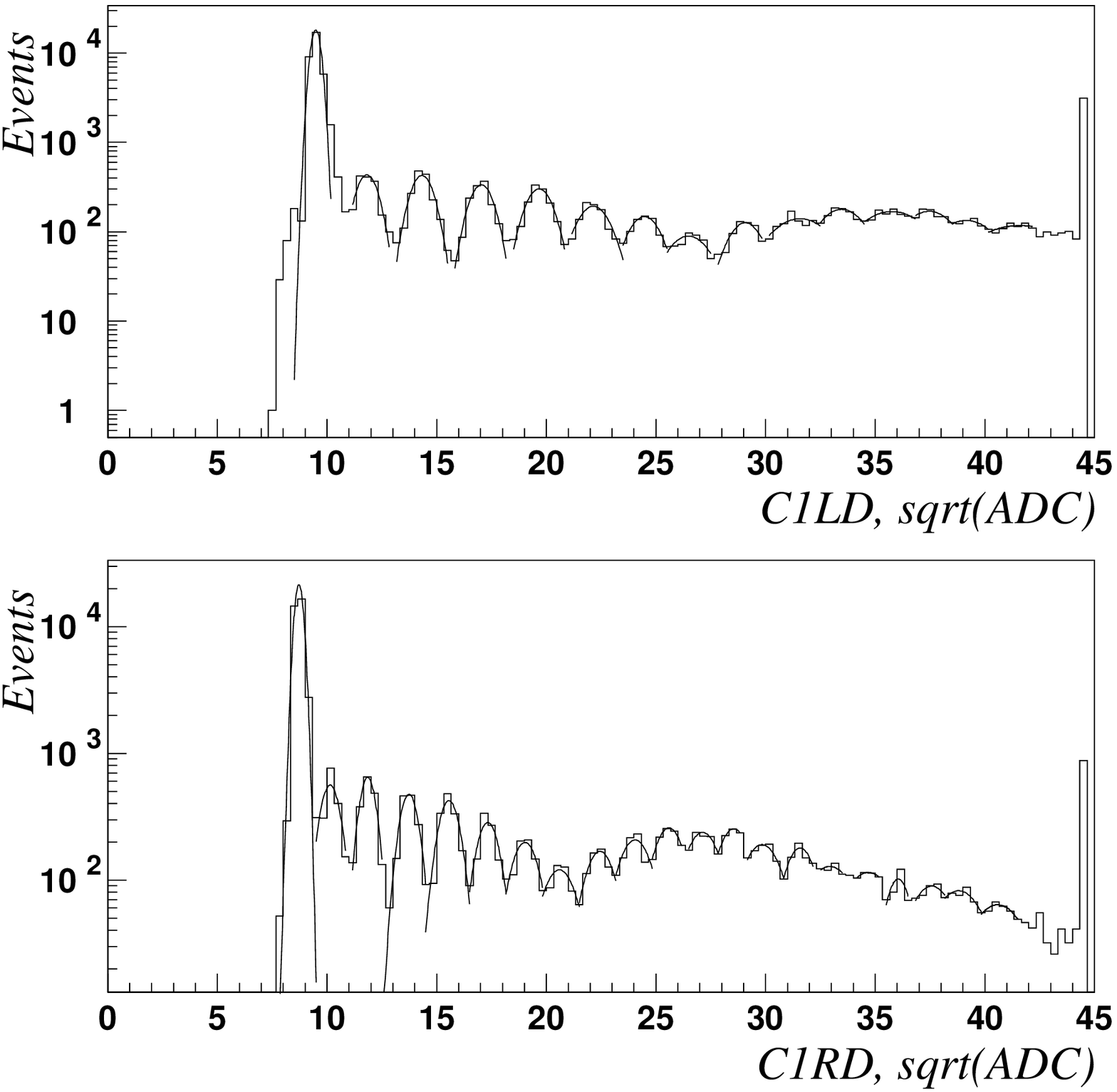}
\end{minipage}
  \caption{Charge peaks observed with the C1 scintillator, using left
  and right anodes (``C1LA'' and ``C1RA'', left panel) and dynodes
  (``C1LD'' and ``C1RD''right panel, where the first peak is He).  The
  3 cm wide beam crossed the counter in the center and the gain ratio
  of the two sides was 3:2.}
\label{peaks}
\end{figure}

\begin{figure}[t]
\begin{minipage}{0.5\textwidth}
\includegraphics[height=0.9\linewidth]{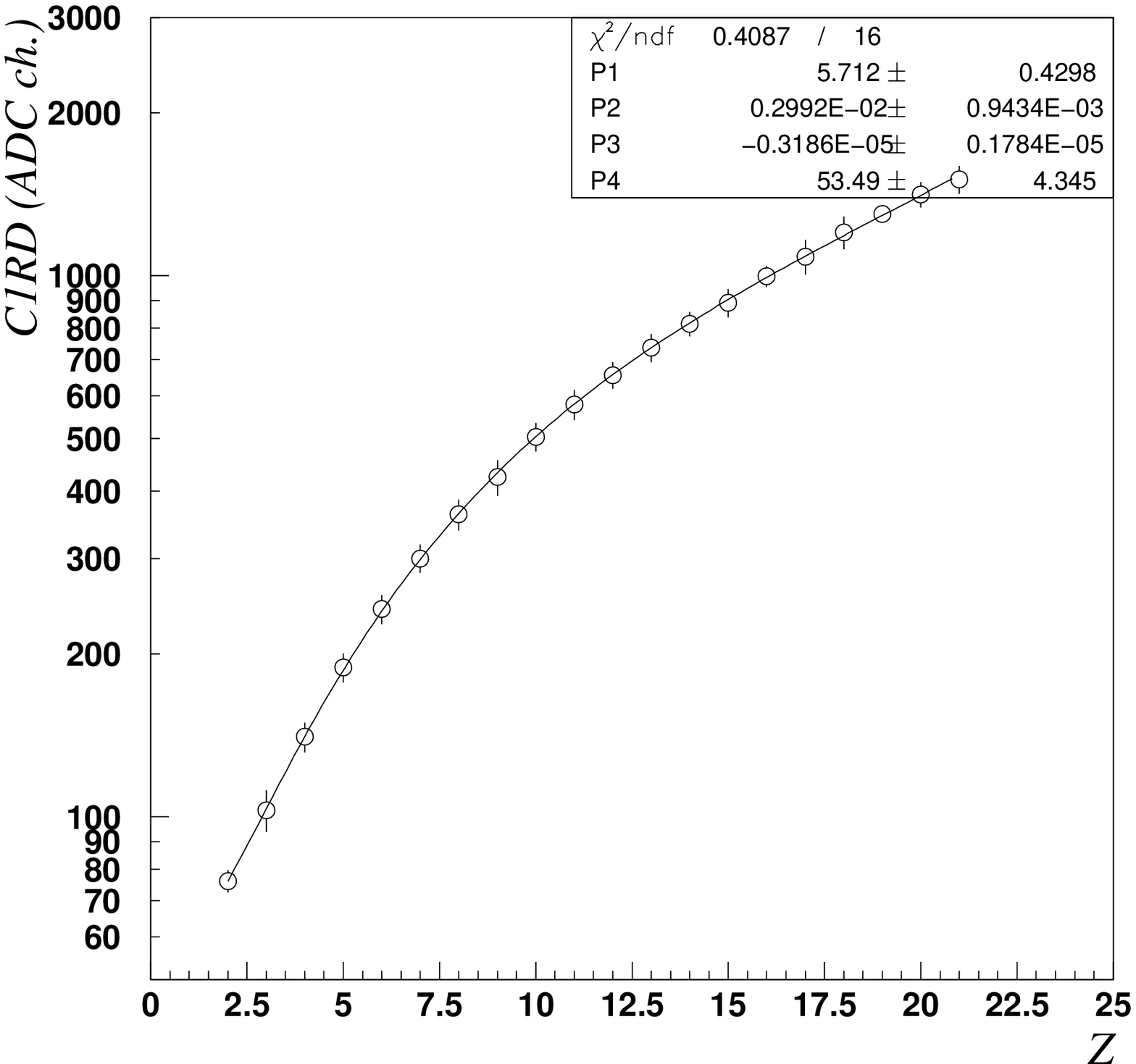}
\end{minipage}%
\begin{minipage}{0.5\textwidth}
~\hfill
\includegraphics[height=0.9\linewidth]{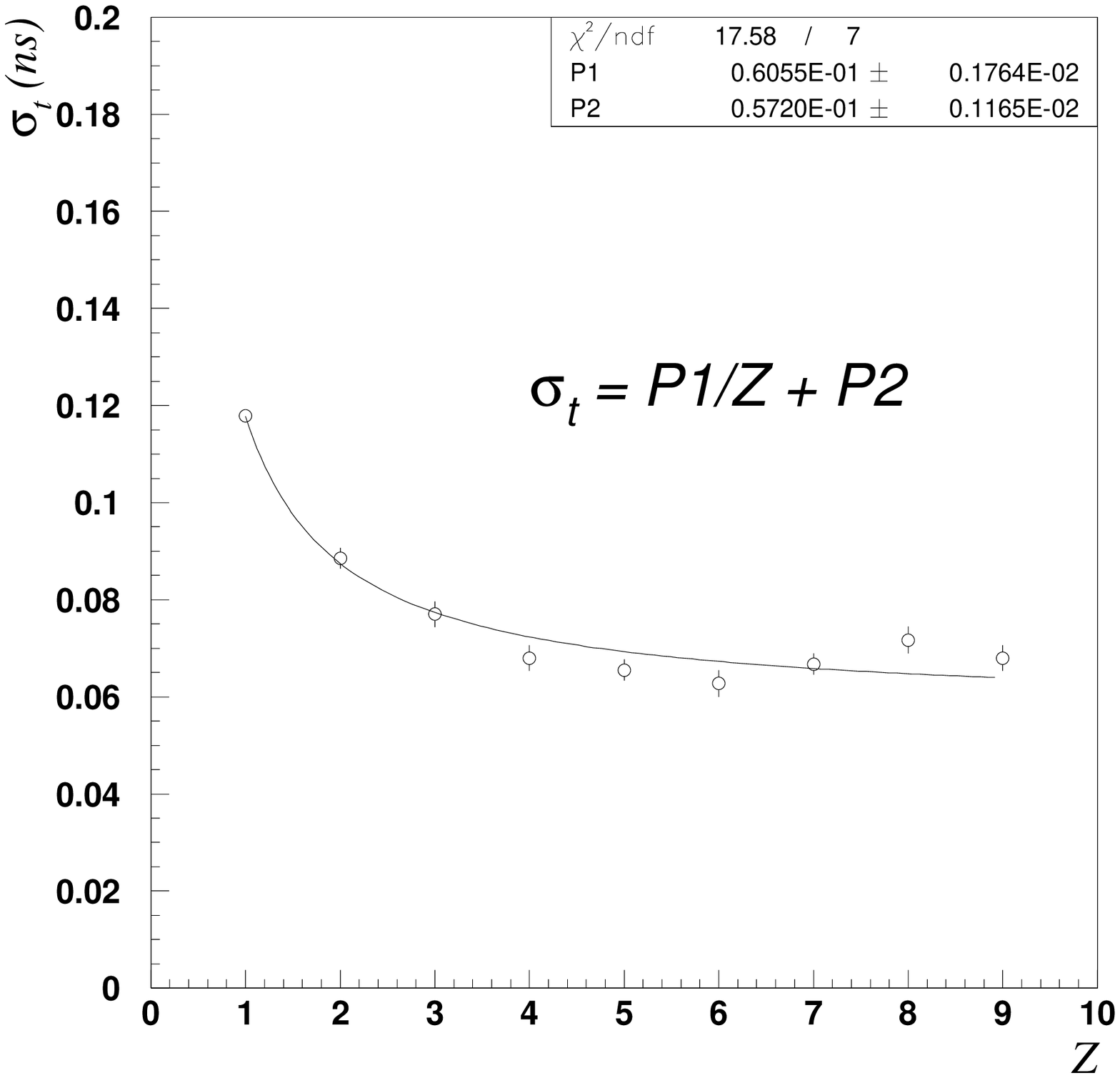}
\end{minipage}
  \caption{Birks' law fit of the dynode peaks (left panel), and
  average time of flight resolution as function of the charge (right
  panel).}\label{bt}
\end{figure}

 The square root of the anode and dynode signals (in arbitrary units)
 of the first scintillator during runs 105--114 is shown in
 figure~\ref{peaks}~ The dynode peaks were fitted with gaussians and
 then the second order Birk's law:
\[
  Q = \frac{A \frac{\mathrm{d}E}{\mathrm{d}x}}{1 
        + B \frac{\mathrm{d}E}{\mathrm{d}x}
        + C \left( \frac{\mathrm{d}E}{\mathrm{d}x} \right)^2}
    = \frac{p_1 Z^2}{1 + p_2 Z^2 + p_3 Z^4}
\]
 (where the constants $A, B, C$ are ``adsorbed'' by $p_1, p_2, p_3$)
 was used to obtain the best estimate of the particle atomic number
 $Z$ from the measured charge $Q = \mathrm{ADC} - p_4$ ($p_4$ is the
 fitted pedestal position), as shown in the left panel of
 figure~\ref{bt}~ The combined TOF and RICH charge measurement is
 shown in ref.~[5].
 
 Finally, the resolution of the measured time of flight between each
 pair of counters can be evaluated as function of the particle charge,
 as shown in the right panel of figure~\ref{bt}~ The time resolution
 improves with increasing charge due to the higher photostatistics,
 until it reaches a level dominated by the electronic noise [3].  With
 the standard electronics used at CERN, this level was $57 \pm 1$ ps,
 while the maximum jitter allowed for the AMS-02 TOF front-end boards
 is 50 ps.

~
\newline
 We wish to thank the organizations and individuals acknowledged in
 ref.~[2].

\section{References}

\re
1.  Aguilar M. et al.,
 Physics Reports, vol. 366/6 (2002), 331-404;
\re
2.  Gentile S.,
 ``The Alpha Magnetic Spectrometer on the International Space Station'',
 these proceedings;
\re
3.  Alvisi D. et al.,
 NIM A 437 (1999) 212-221;
\re
4.  Brocco L. et al.,
 Proc. of the 27th ICRC (2001), 2193-2196;
\re
5.  Bu\'enerd M., 
 ``The AMS RICH detector'',
 these proceedings.

\endofpaper
\end{document}